\documentclass[prd,twocolumn,floatfix,amsmath,nofootinbib,amssymb,floatfix]{revtex4}
\usepackage{graphicx,color,dcolumn,booktabs,bm}
\usepackage{longtable,lscape}
\usepackage{txfonts}
\usepackage{overpic}
\usepackage{amssymb}
\usepackage{indentfirst}
\usepackage{feynmf}   
\usepackage{slashed}  
\usepackage{cases}
\usepackage{color}
\usepackage{multirow}
\usepackage{epstopdf}
\usepackage{enumerate}
\usepackage{graphicx,color,dcolumn,booktabs,bm}
\usepackage[colorlinks, citecolor=blue,anchorcolor=red,menucolor=red, linkcolor=red,filecolor=red,urlcolor=blue,frenchlinks=red]{hyperref}

\begin{document}

\title{Toward charged $Z_{cs}(3985)$ structure under a reflection mechanism}
\author{Jun-Zhang Wang$^{1,2}$}\email{wangjzh2012@lzu.edu.cn}
\author{Qin-Song Zhou$^{1,2}$}\email{zhouqs13@lzu.edu.cn}
\author{Xiang Liu$^{1,2}$\footnote{Corresponding author}}\email{xiangliu@lzu.edu.cn}
\author{Takayuki Matsuki$^{3}$}\email{matsuki@tokyo-kasei.ac.jp}
\affiliation{$^1$School of Physical Science and Technology, Lanzhou University, Lanzhou 730000, China\\
$^2$Research Center for Hadron and CSR Physics, Lanzhou University $\&$ Institute of Modern Physics of CAS, Lanzhou 730000, China\\
$^3$Tokyo Kasei University, 1-18-1 Kaga, Itabashi, Tokyo 173-8602, Japan
}

\date{\today}

\begin{abstract}
Very recently, the BESIII collaboration reported a charged hidden-charm structure with strangeness in the recoil mass of $K^+$ of a process $e^+e^-\to D_s^{*-}D^0K^+$ or $D_s^{-}D^{*0}K^+$, which is named as $Z_{cs}(3985)^{-}$. The newly observed charged structure can be treated as a partner structure with strangeness of well-known $Z_{c}(3885)^{-}$ reported in a process $e^+e^-\to D^{*-}D^0\pi^+$. In this work, we propose a reflection picture to understand the nature of $Z_{cs}(3985)$. By performing a combined analysis for the line shape of the recoil mass distribution of $K^+$ at five energy points $\sqrt{s}=4.628, 4.641, 4.661, 4.681, 4.698$ GeV, we find that the $Z_{cs}(3985)$ can be explained as a reflection structure of charmed-strange meson $D_{s2}^{*}(2573)$, which is produced from the open-charm decay of $Y(4660)$ with a $D_s^*$ meson. Furthermore, we predicted the angular distribution of final state $D_s^{*-}$ in process $e^+e^-\to D_s^{*-}D^0K^+$ based on our proposed reaction mechanism, which may be an essential criterion to test the non-resonant nature of $Z_{cs}(3985)$ further.

\end{abstract}

\maketitle


In the past two decades, the researches on exotic $XYZ$ hadronic states have been one of the hottest topics in particle physics \cite{Chen:2016qju,Liu:2019zoy,Guo:2017jvc,Olsen:2017bmm,Brambilla:2019esw}, whose novel properties have brought a great challenge for the understanding of strong interaction. In the $XYZ$ family,  the charged heavy quarkoniumlike $Z_{Q}$ structures by electron-positron annihilation are an exceptional class of exotic structures. Since the first observation of $Z_c(3900)^{\pm}$ in the invariant mass spectrum of $J/\psi \pi^{\pm}$ of a process $e^+e^- \to J/\psi \pi^+ \pi^-$ by BESIII \cite{Ablikim:2013mio}, a series of charged $Z_c$ structures were experimentally discovered ~\cite{Liu:2013dau,Ablikim:2013wzq,Ablikim:2013xfr,Ablikim:2013emm,Ablikim:2017oaf}, whose charged behavior can securely exclude the conventional candidate of heavy quarkonium. Thus, the $Z_c$ structures must imply a novel nature different from mesonic hadrons in the quark model.

Very recently, the BESIII collaboration released measurements of $e^+e^-\to  D_s^{*-}D^0K^+$ or $D_s^{-}D^{*0}K^+$ at $\sqrt{s}=4.681$ GeV, where a charged hidden-charm structure $Z_{cs}(3985)^{-}$ was observed in the vicinity of the threshold of $D_s^{*-}D^0$ ($D_s^{-}D^{*0}$) \cite{1830518}. Surprisingly, this is the first observation of charged heavy quarkoniumlike structures with strangeness, whose discovery could provide some unique hints to uncover the secrets of charged exotic $Z$ structures. From the production mode, it can be easily concluded that $Z_{cs}(3985)^{-}$ should be a partner structure with strangeness of well-known $Z_{c}(3885)^{-}$ reported in $e^+e^-\to  D^{*-}D^0\pi^+$ \cite{Ablikim:2013xfr}, which means that $Z_{c}(3885)^{-}$ and $Z_{cs}(3985)^{-}$ are governed by a similar production mechanism.
In the past theoretical opinions on $Z_c(3885)^{\pm}$, the tetraquark explanation with $c\bar{c}u\bar{d}$ is the most popular and other interpretations include triangle singularity, hadronic molecular state, etc. (see review articles \cite{Chen:2016qju,Liu:2019zoy,Guo:2017jvc,Olsen:2017bmm,Brambilla:2019esw} for more details). Furthermore, based on the explanations of $Z_c$ structures, several theoretical models have also been extended to predict the properties of $Z_{cs}$ \cite{Lee:2008uy,Ferretti:2020ewe,Dias:2013qga,Chen:2013wca}. These explanations basically cover various exotic hadron configurations.

Recently, the Lanzhou group proposed a new universal non-resonant explanation to $Z_c(3885)$ and $Z_c(4025)$ \cite{Wang:2020axi}, whose line shapes on the respective invariant mass distribution are found to be well described by the reflection peak from the $P$-wave charmed meson $D_1(2420)$ as an intermediate resonance involved in a process $e^+e^-\to  D^{(*)}D_1(2420)  \to D^{(*)}\bar{D}^{*}\pi$. 
Of course, if this reflection mechanism is really the key to the nature of charmoniumlike structures $Z_c(3885)$ and $Z_c(4025)$, we expect that it should also play a crucial role in a process $e^+e^-\to  D_s^{*-}D^0K^+$ or $D_s^{-}D^{*0}K^+$. Based on this idea, in this work, we will explore the possibility of newly observed $Z_{cs}(3985)^{-}$ in the proposed reflection mechanism. Our results indicate that $Z_{cs}(3985)^{-}$ can be naturally explained as a reflection structure from a charmed-strange meson $D_{s2}^*(2573)$. This finding not only provides a view to decode the inner structure of $Z_{cs}(3985)^{-}$ but also strongly demonstrates the universality of the reflection mechanism in describing the charged charmoniumlike structures. At the same time, we also discuss the implication of $Z_{c(s)}$ structures having a connection with charmoniumlike $Y$ states. In the following, we will illustrate them in detail.


At present, the observed charged hidden-charm structures in  electron-positron annihilation including $Z_c(3885)^{\pm}$, $Z_c(3900)^{\pm}$, $Z_c(4025)^{\pm}$, etc., come from a hadronic decay of the vector charmoniumlike state $Y(4220)$ \cite{Liu:2013dau,Ablikim:2013wzq,Ablikim:2013xfr,Ablikim:2013emm}. It is no doubt that a full understanding of $Y(4220)$ can greatly limit different candidate interpretations of these $Z_c$ structures.  In Ref. ~\cite{Wang:2019mhs}, the Lanzhou group adopted an unquenched quark model to revisit the mass spectrum of $J/\psi$ family, where $Y(4220)$ and $Y(4360)$ are found to be explained very well as the mixing eigenstates between pure charmonium states $\psi(4S)$ and $\psi(3D)$. From the charmonium view of $Y(4220)$, it will have a strong coupling with the $S$-wave two-body open charm channel $D\bar{D}_1(2420)$ \cite{Wang:2019mhs}, which  is indeed supported by a recent measurement of BESIII \cite{Ablikim:2018vxx}. Furthermore, the $D_1(2420)$ will dominantly decay into $D^* \pi$ and then its contribution can cause a reflection peak near the threshold on the invariant mass spectrum of $DD^*$ in $e^+e^-\to D\bar{D}^{*}\pi$, which can just explain a $Z_c(3885)$ structure \cite{Wang:2020axi}.

Similarly, for the partner structure with strangeness of $Z_c(3885)^{-}$, the BESIII measurements indicated that a clear signal of $Z_{cs}(3985)^{-}$ only appears at center of mass (CM) energy of 4.681 GeV.  In the vicinity of this energy point, there also exists a vector charmoniumlike state $Y(4660)$, which has been reported in both processes $e^+e^- \to \psi(3686)\pi^+\pi^-$ \cite{Wang:2007ea} and $e^+e^- \to \Lambda_c\Lambda_c$ \cite{Pakhlova:2008vn} by Belle. In Ref. ~\cite{Wang:2020prx}, based on the same unquenched picture, it was found that $Y(4660)$ can also be accommodated in the $J/\psi$ family and is treated as a good candidate of a higher charmonium eigenstate by $6S$-$5D$ mixture. Thus, along the line of $Y(4220)$ and $Y(4660)$ as charmonia in a unified theoretical framework \cite{Wang:2019mhs,Wang:2020prx}, we believe that there must exist similarity between the production mechanisms of $Z_{cs}(3985)^{-}$ and  $Z_c(3885)^{-}$.
We can conclude that the contributions from two-body open charm channels $D^{(*)}\bar{D}^{**}$ or $D_s^{(*)}\bar{D}_s^{**}$ by the strong decay of $Y(4660)$ should be dominant for the three-body open charm processes $e^+e^-\to  D_s^{*-}D^0K^+$ and $e^+e^-\to D_s^{-}D^{*0}K^+$, where  $\bar{D}^{**}$ and $\bar{D}_s^{**}$ stand for higher excited charmed and charmed-strange mesons, respectively.  Here, the dynamical and kinematical behaviors owing to some special $D^{**}$ or $D_s^{**}$ mesons may produce the similar reflection phenomenon on invariant mass distribution of $D_s^{*-}D^0$ or $D_s^{-}D^{*0}$, which will provide a reflection perspective to understand the newly observed $Z_{cs}(3985)$ structure.

\begin{figure}[b]
	\includegraphics[width=7.5cm,keepaspectratio]{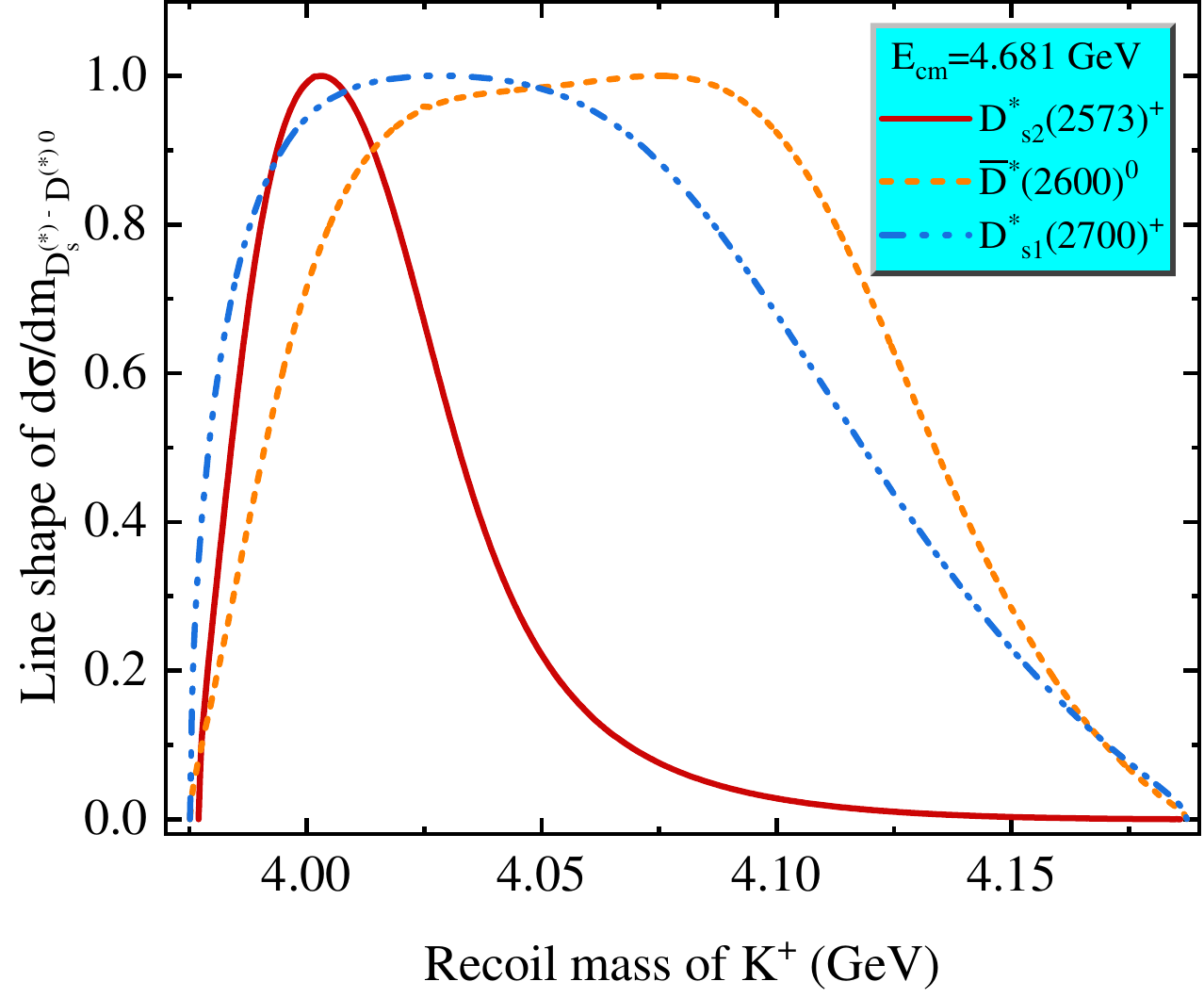}
	\caption{Line shapes of recoil mass spectra of $K^{+}$ for $e^+e^-\to D^{*}_{s2}(2573)^{+}D_{s}^{*-}\to(D^0 K^+)D_{s}^{*-}$, $e^+e^-\to \bar{D}^{*}(2600)^{0}D^{*0}\to(D_{s}^{-} K^+)D^{*0}$, and $e^+e^-\to D^{*}_{s1}(2700)^{+}D_{s}^{-}\to(D^{*0} K^+)D_{s}^{-}$, respectively.}\label{linesshape}
\end{figure}

Reflection phenomenon means that in a general scattering cascade process $AB \to CD \to C(EF)$, the intermediate resonance $D$, which can be directly seen in the invariant mass spectrum of $EF$ as a Breit-Wigner distribution, can be reflected into the other two invariant mass distributions of $CE$ or $CF$. Under some special kinematical behaviors, the intermediate resonance can show an obvious reflection peak near the threshold of invariant mass spectrum. The concrete details for the reflection mechanism can be seen in Ref. \cite{Wang:2020dmv}, which gives a general criterion to identify whether reflection peak near threshold can be formed, i.e., $m_{C}+m_{D}\geq \sqrt{s}$, where $\sqrt{s}$ is the CM energy of system. Here, the case of $m_{C}+m_{D}> \sqrt{s}$ means intermediate resonance $C$ is always off-shell while $m_{C}+m_{D}= \sqrt{s}$ can cause a single on-shell pole in the phase space, which just corresponds to a clear reflection peak near threshold.  Based on this guideline, we initially select three candidate processes of $e^+e^-\to D^{*0}\bar{D}^{*}(2600)^0 \to D^{*0}(D_s^-K^+)$, $e^+e^-\to D_s^{*-}D_{s2}^*(2573)^+ \to D_s^{*-}(D^{0}K^+)$ and $e^+e^-\to D_{s}^-D_{s1}^*(2700)^+ \to D_{s}^-(D^{*0}K^+)$ by setting $\sqrt{s}=$4.681 GeV.

In the following, we will study the reflective line shape from intermediate resonance $\bar{D}^{*}(2600)^0$ \cite{delAmoSanchez:2010vq}, $D_{s2}^*(2573)^+$ \cite{Kubota:1994gn}, and $D_{s1}^*(2700)^+$ \cite{Brodzicka:2007aa} at $\sqrt{s}=$4.681 GeV. Here, the effective Lagrangian approach is adopted, where the related Lagrangian densities are listed below  ~\cite{Bauer:1975bv,Bauer:1975bw,Casalbuoni:1996pg,Chen:2011xk,Liu:2020ruo}
\begin{eqnarray}
\mathcal{L}_{\gamma Y}&=&\frac{-em_Y^2}{f_Y}Y_{\mu}A^{\mu},\nonumber\\
\mathcal{L}_{D^*{D}^{*\prime}Y}&=&-ig_{D^*{D}^{*\prime}Y}Y_{\mu}(D^{*\nu\dag}\partial^{\nu}D^{*\prime}_{\mu} \nonumber \\
&&-\partial^{\nu}D^{*\dag}_{\mu}D_{\nu}^{*\prime}-D^{*\nu\dag}\overleftrightarrow{\partial}_{\mu}D^{*\prime}_{\nu})+c.c., \nonumber \\
\mathcal{L}_{KD_sD^{*\prime}}&=&-ig_{KD_sD^{*\prime}}(D_s^{\dag}(\partial_{\mu}K)D^{*\prime\mu}-D^{*\prime\mu\dag}(\partial_{\mu}K)D_s), \nonumber \\
\mathcal{L}_{D_{s2}{D}_s^*Y}&=&g_{D_{s2}D_s^*Y}Y_{\mu}(D_{s2}^{\mu\nu\dag}D^*_{s\nu}+D^{*\dag}_{s\nu}D_{s2}^{\mu\nu}), \nonumber \\
\mathcal{L}_{KD{D}_{s2}}&=&g_{KDD_{s2}}(D_{s2}^{\mu\nu\dag}(\partial_{\mu}\partial_{\nu}K)D+D^{\dag}(\partial_{\mu}\partial_{\nu}K)D_{s2}^{\mu\nu}),  \nonumber \\
\mathcal{L}_{D_sD_{s1}^*Y}&=&-g_{D_sD_{s1}^*Y}\varepsilon^{\mu\nu\alpha\beta}\partial_{\mu}Y_{\nu}D_s^{\dag}\overleftrightarrow{\partial}_{\beta}D^*_{s1\alpha}+c.c.
, \nonumber \\
\mathcal{L}_{KD^*D_{s1}^*}&=&g_{KD^*D_{s1}^*}\varepsilon_{\mu\nu\alpha\beta}D^{*\mu\dag}\partial^{\nu}K\overleftrightarrow{\partial}^{\alpha}D_{s1}^{*\beta}+c.c..
\end{eqnarray}
Here, $A^{\mu}$ and $Y^{\mu}$ are photon and $Y(4660)$ fields, respectively, and $g_{ABC}$ stands for the corresponding coupling constant.
Based on the above coupling vertices, the line shapes of the differential cross sections of $e^+e^-\to D^{*0}\bar{D}^{*}(2600)^0 \to D^{*0}(D_s^-K^+)$, $e^+e^-\to D_s^{*-}D_{s2}^*(2573)^+ \to D_s^{*-}(D^{0}K^+)$, and $e^+e^-\to D_{s}^-D_{s1}^*(2700)^+ \to D_{s}^-(D^{*0}K^+)$ vs. the invariant mass $m_{D_{s}^{(*)-}D^{(*)0}}$ can be directly shown in Fig. \ref{linesshape}. It can be seen that the reflective distributions both from $\bar{D}^{*}(2600)^0$ and $D_{s1}^*(2700)^+$ present the non-peaking shape, which cannot explain the $Z_{cs}(3985)^{-}$ structure. By contrast, there exists an obvious reflective peak near threshold in contribution involving the channel $D_s^{*-}D_{s2}^*(2573)^+$. Thus, after a comprehensive analysis, we conjecture that  $D_{s2}^*(2573)^+$ \cite{Kubota:1994gn} should be a unique intermediate $D_{(s)}^{**}$ candidate, which could provide a reflection explanation to $Z_{cs}(3985)^{-}$. At the same time, a recent study also indicates that there exists a strong coupling between charmed-strange meson $D_{s}^{*}$ and $D^{*}_{s2}(2573)$ \cite{Wang:2020dya}.

\begin{figure}[t]
  \centering
  \begin{tabular}{cc}
  \includegraphics[width=120pt]{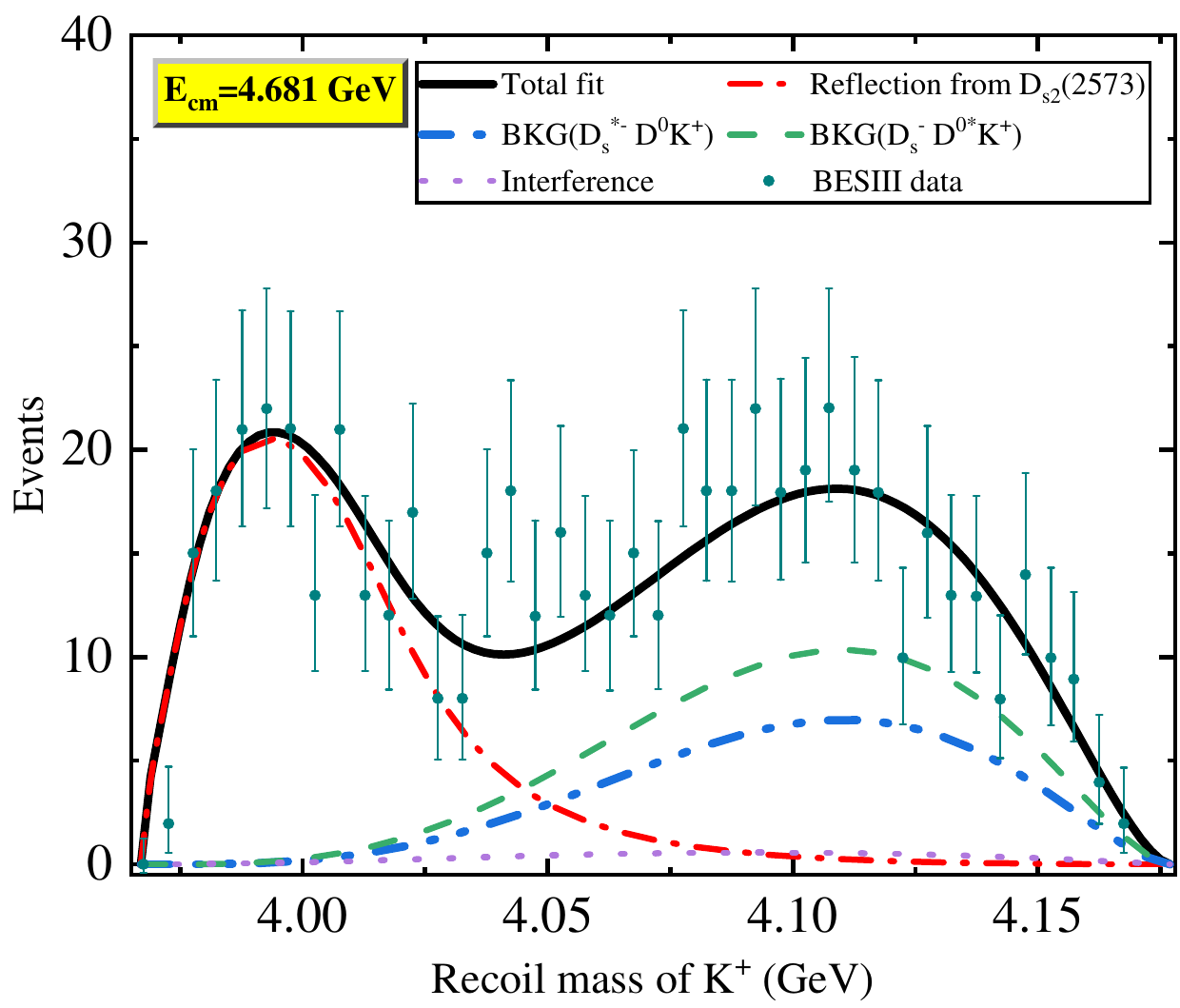} & \includegraphics[width=117pt]{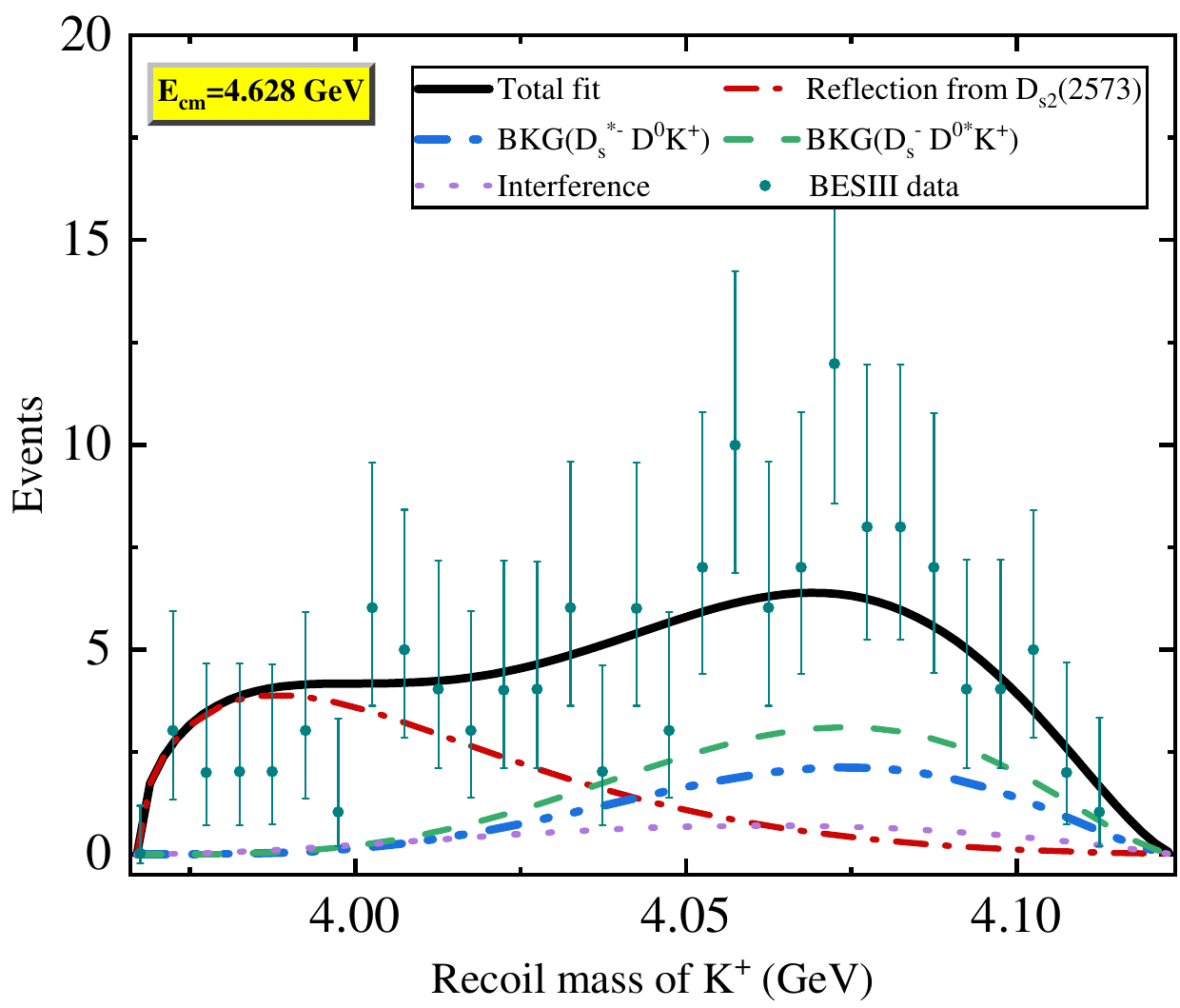} \\
  \includegraphics[width=120pt]{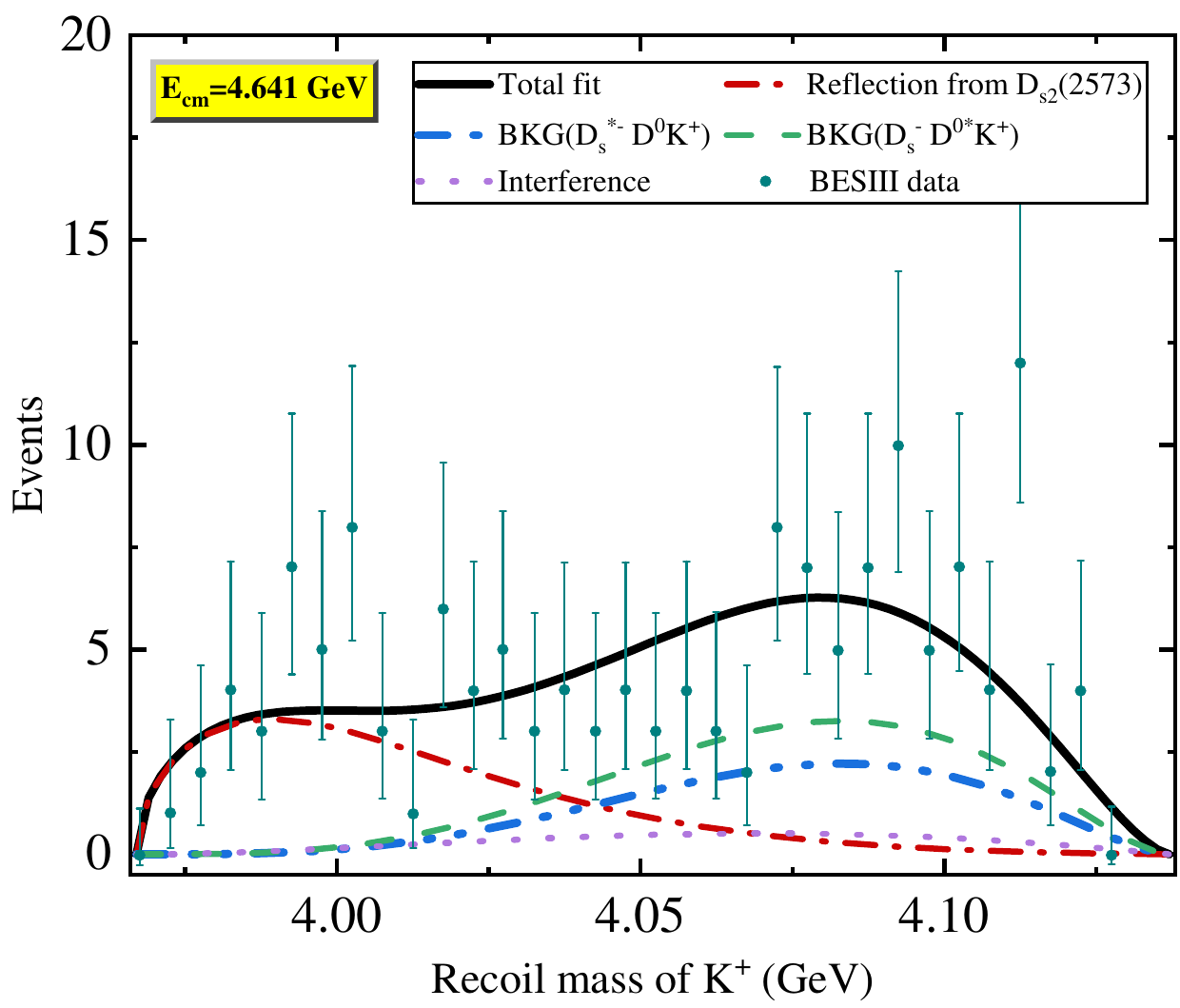} & \includegraphics[width=125pt]{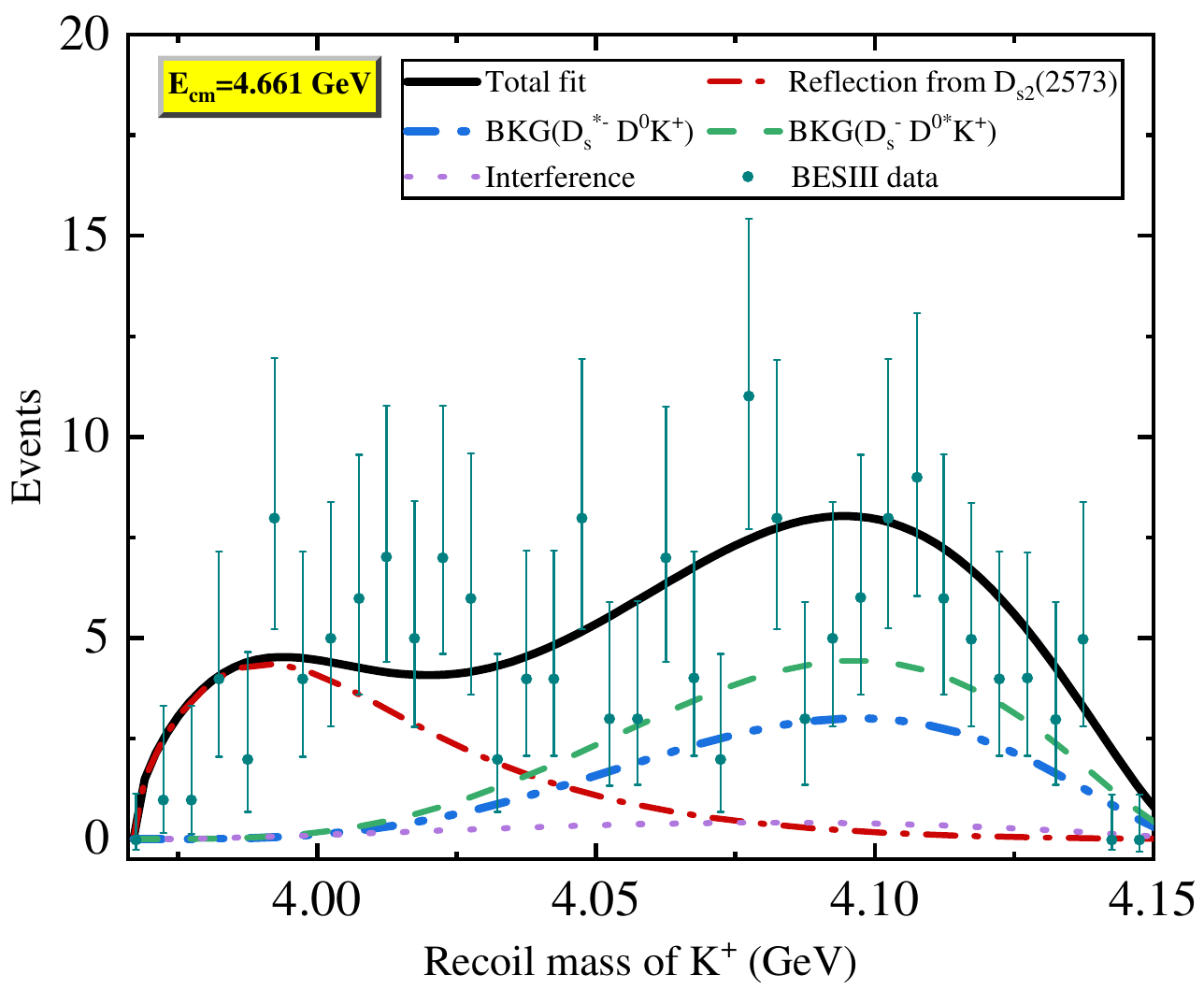}  \\
   \includegraphics[width=120pt]{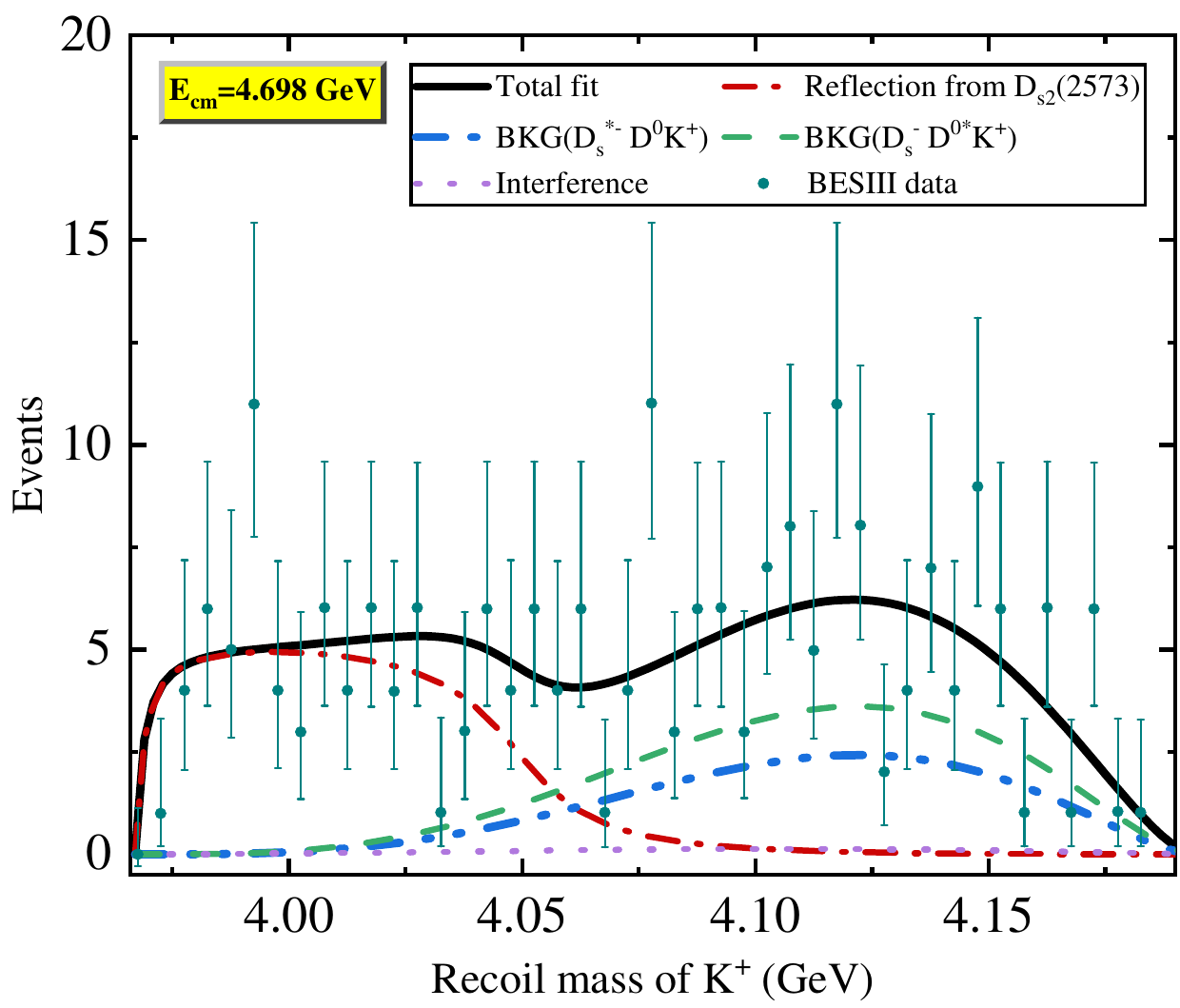} & \includegraphics[width=116pt]{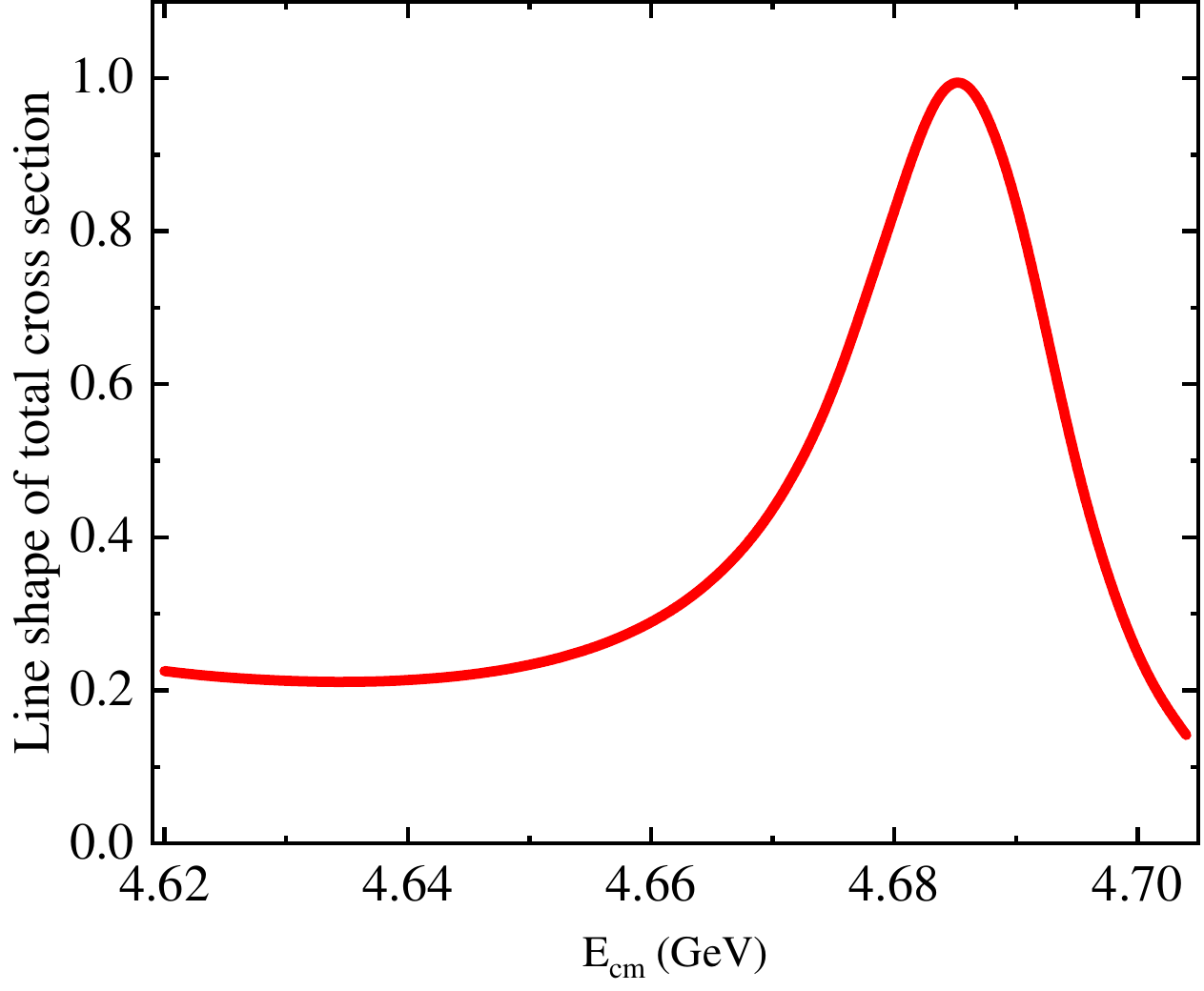}  \\
  \end{tabular}
 \caption{ The fit to the experimental data of $e^+e^-\to D_{s}^- D^{*0}K^{+}$ or $D_{s}^{*-}D^{0}K^+$ by line shapes on recoil mass spectrum of $K^{+}$ at five CM energy points. Additionally, the theoretical line shape of the total cross section based on above fit is also performed. Here, only the reflection from charmed-strange meson $D_{s2}^{*}(2573)$ and normal nonpeaking contributions are included. \label{Fitresult} }
\end{figure}

To verify the above idea, we can make a combined analysis for experimental recoil mass spectra of $K^+$ at five CM energy points in our theoretical framework. Here, we mainly consider three contributions, i.e., the reflection from $D_{s2}^*(2573)^+$, and two non-peaking backgrounds with final states $D_{s}^-D^{*0}K^+$ and $D_{s}^{*-}D^{0}K^+$. Referring to the treatments in Ref. \cite{Wang:2020axi}, their scattering amplitudes can be written as
\begin{eqnarray}
\mathcal{M}^{D_{s2}^*(2573)}&=&\frac{-\mathcal{A}^{e^+e^- \to Y(4660)}_{\rho} g_{D_{s2}D_s^*Y}g_{KDD_{s2}}\epsilon_{D_s^*}^{\lambda *}G_{\rho\lambda\alpha\beta}p_2^{\alpha}p_2^{\beta}}{(p_1-p_4)^2-m_{D_{s2}}^2+im_{D_{s2}}\Gamma_{D_{s2}}}, \nonumber  \\
\mathcal{M}_{D_{s}D^{*}K}^{Nonpeak}&=&\frac{\mathcal{A}^{e^+e^- \to Y(4660)}_{\rho} g_{YD_s{D}^*K}\epsilon_{D^*}^{\rho *}}{(m_{D_{s}D^{*}}-m_{D_{s}}-m_{D^{*}})^{-a_1}(\sqrt{s}-m_{K}-m_{D_{s}D^{*}})^{-b_1}},  \nonumber  \\
\mathcal{M}_{D_{s}^{*}DK}^{Nonpeak}&=&\frac{\mathcal{A}^{e^+e^- \to Y(4660)}_{\rho} g_{YD_s^*{D}K}\epsilon_{D_s^*}^{\rho *}}{(m_{D_{s}^*D}-m_{D_{s}^*}-m_{D})^{-a_2}(\sqrt{s}-m_{K}-m_{D_{s}^*D})^{-b_2}} \nonumber
\end{eqnarray}
with
\begin{eqnarray}
\mathcal{A}^{e^+e^- \to Y(4660)}_{\rho}=\bar{v}(k_2)e\gamma_{\mu}u(k_1)\frac{g^{\mu\nu}em_{Y}^2(g_{\nu\rho}-p_{1\nu}p_{1\rho}/m_Y^2)}{sf_{Y}(s-m_Y^2+im_Y\Gamma_Y)},
\end{eqnarray}
respectively,
where $G_{\rho\lambda\alpha\beta}=\frac{1}{2}(\tilde{g}_{\rho\alpha}\tilde{g}_{\lambda\beta}+\tilde{g}_{\rho\beta}\tilde{g}_{\lambda\alpha})-\frac{1}{3}\tilde{g}_{\rho\lambda}\tilde{g}_{\alpha\beta}$ with $\tilde{g}_{\rho\alpha}=-g_{\rho\alpha}+(p_{1}-p_{4})_\rho(p_{1}-p_{4})_\alpha/m_{D_{s2}}^2$. Here, the four-momentum $p_1$, $p_2$, $p_3$, $p_4$ are pointed to $Y(4660)$, $K$, $D_{(s)}$, and $D_{(s)}^*$, respectively.  The factor $(m_{D_{s}^{(*)}D^{(*)}}-m_{D_{s}^{(*)}}-m_{D^{(*)}})^{-a_i}(\sqrt{s}-m_{K}-m_{D_{s}^{(*)}D^{(*)}})^{-b_i}$ is introduced to phenomenologically absorb non-peaking contributions from other intermediate charmed and charmed-strange mesons \cite{Ablikim:2013xfr}.
Then, the differential cross section for the recoil spectra of $K^+$ can be expressed by
\begin{eqnarray}
\frac{d\sigma}{dm_{R}}=\frac{\int\left|\bf{p_2}\right|\left|\bf {p_3}^*\right|\overline{\left|\mathcal{M}_1\right|^2}d\Omega_{2}d\Omega_{3}^*+\int\left|\bf{p^{\prime}_2}\right|\left|\bf {p^{\prime}_3}^*\right|\overline{\left|\mathcal{M}_2\right|^2}d\Omega^{\prime}_{2}d\Omega_{3}^{\prime*}}{32(2\pi)^5(k_1\cdot k_2)\sqrt{s}} \nonumber
\end{eqnarray}
with
\begin{eqnarray}
\mathcal{M}_1&=&\mathcal{M}_{D_{s}^{*}DK}^{Nonpeak}\cdot\mathcal{F}_{Y}(s)+e^{i\phi}\mathcal{M}^{D_{s2}^*(2573)}\cdot\mathcal{F}^{\prime}_{Y}(s), \nonumber \\
 \mathcal{M}_2&=&\mathcal{M}_{D_{s}D^{*}K}^{Nonpeak}\cdot\mathcal{F}_{Y}(s),
\end{eqnarray}
where the overline above an amplitude stands for the average over spin of initial states and the sum over spin of final states, and the symbol star above physical quantities means that they are quantities in the rest frame of the $D_s^{(*)}D^{(*)}$ system. 
{The form factor $\mathcal{F}^{(\prime)}_{Y}(s)=e^{-c^{(\prime)}(\sqrt{s}-m_{Y})}$ is introduced to modify the Breit-Wigner distribution of the $Y(4660)$ state, which can balance the relative size of total cross sections at different energies. As for the adopted form factor, we follow a previous work of our research group \cite{Huang:2019agb} , where the expression $e^{-c(\sqrt{s}-m_{Y})}$ has been successfully applied to the description of cross sections. Here, it is worth emphasizing that the introduction of a form factor $F_{Y}(s)$ only affects the line shape of the $Y$ state in the total cross section without changing our fit for the line shape of $Z_{cs}(3985)$. }

With the above preparations, we can perform a combined fit to recoil mass spectra of $K^+$ at five CM energy points by BESIII, which are presented in Fig. \ref{Fitresult}. In addition, the relevant fitting parameters are summarized in Table \ref{parameters}.  It can be clearly seen that the signal of $Z_{cs}(3885)^{-}$ at $\sqrt{s}=4.681$ GeV can be described well by just the reflection peak from $D_{s2}^*(2573)^+$. In fact, what makes the $D_{s2}^*(2573)$ so special is that it is quite a coincidence that $m_{D_{s2}^*(2573)}+m_{D_{s}^*}=4.680$ GeV is almost equal to the measured CM energy point of $\sqrt{s}=4.681$ GeV. According to the research findings in Ref. \cite{Wang:2020dmv}, the above situation just satisfies the critical relation of producing the near-threshold reflection peak, which can correspond to a most prominent peak line shape. This can naturally explain why the BESIII experiment observed an obvious charged near-threshold structure only at  $\sqrt{s}=4.681$ GeV as seen in Fig. \ref{Fitresult}. One can see that the reflection from $D_{s2}^*(2573)^+$ behaves an unclear peak at $\sqrt{s}=4.628, 4.641, 4.661$ GeV and non-peaking platform shape at $\sqrt{s}=4.698$ GeV. Thus, the $\sqrt{s}=4.681$ GeV is a very special energy point to stimulate the appearance of a reflection peak phenomenon. These evidences should provide a strong support to our proposed reflection explanation to $Z_{cs}(3985)$.

\begin{table*}
\centering
\caption{ The parameters for fitting the experimental line shape on the recoil mass spectra of $K^+$ of $e^+e^-\to D_s^{*-}D^0K^+ +D_s^{-}D^{*0}K^+$.}\label{parameters}
\setlength{\tabcolsep}{1.3mm}{
\begin{tabular}{ccccccccccccc}
\toprule[1.0pt]
Parameters & $\mid \frac{g_{D_{s2}D_s^*Y} g_{KDD_{s2}}}{g_{YD_s^*{D}K}}\mid$&$\mid\frac{g_{YD_s{D}^*K}}{g_{YD_s^*{D}K}}\mid$&$\phi$ (rad)&$c$~(GeV$^{-1}$)&$c^{\prime}$~(GeV$^{-1}$)&$a_1$&$b_1$ &$a_2$&$b_2$&$\Gamma_{Y}$~(GeV) &$m_{Y}$~(GeV) & $\chi^2/d.o.f.$ \\
\midrule[0.8pt]
Value & 0.0279 & 1.29 &0.288&32.1&48.0&1.53&0.562 &1.54&0.574 &0.022&4.687 & 0.762\\
Error($\pm$) &0.0201 & 0.93 & 0.840 & 1.3 & 1.4 & 0.07 & 0.096 & 0.03 & 0.065 & 0.001 & 0.001 & $-$ \\
\bottomrule[1.0pt]
\end{tabular}}
\end{table*}

After carrying out a combined analysis to the recoil mass spectra of $K^+$, utilizing the fitting parameters in Table \ref{parameters}, we can directly calculate the line shape of total cross section of $e^+e^-\to Y(4660) \to D_s^{*-}D^0K^+ +D_s^{-}D^{*0}K^+$, which are depicted in Fig. \ref{Fitresult}. It can be seen that our theoretical description for the total cross section shows that the signal of $Y(4660)$ is very evident. Here, it is worth mentioning that the fitting resonance parameters of $Y(4660)$ are found to be $m_{Y}=4687 \pm1$ MeV and $\Gamma_{Y}=22\pm1$ MeV, respectively, which are in good agreement with our previous theoretical estimate of a mass of 4675 MeV and a width of 30 MeV for $Y(4660)$ in Ref.  \cite{Wang:2020prx}. Thus, from this point, the BESIII data in three-body open-charm processes of $e^+e^-\to  D_s^{(*)-}D^{(*)0}K^+$ also support the charmonium nature of $Y(4660)$. Hence, our present studies together with a series of previous works \cite{Wang:2020axi,Wang:2019mhs,Wang:2020prx} have completely established a natural connection between $Y$ states and $Z_{c(s)}$ structures in $XYZ$ family, which can continue to be tested in more precise experimental data in the future.

\begin{figure}[ht]
	\includegraphics[width=7.5cm,keepaspectratio]{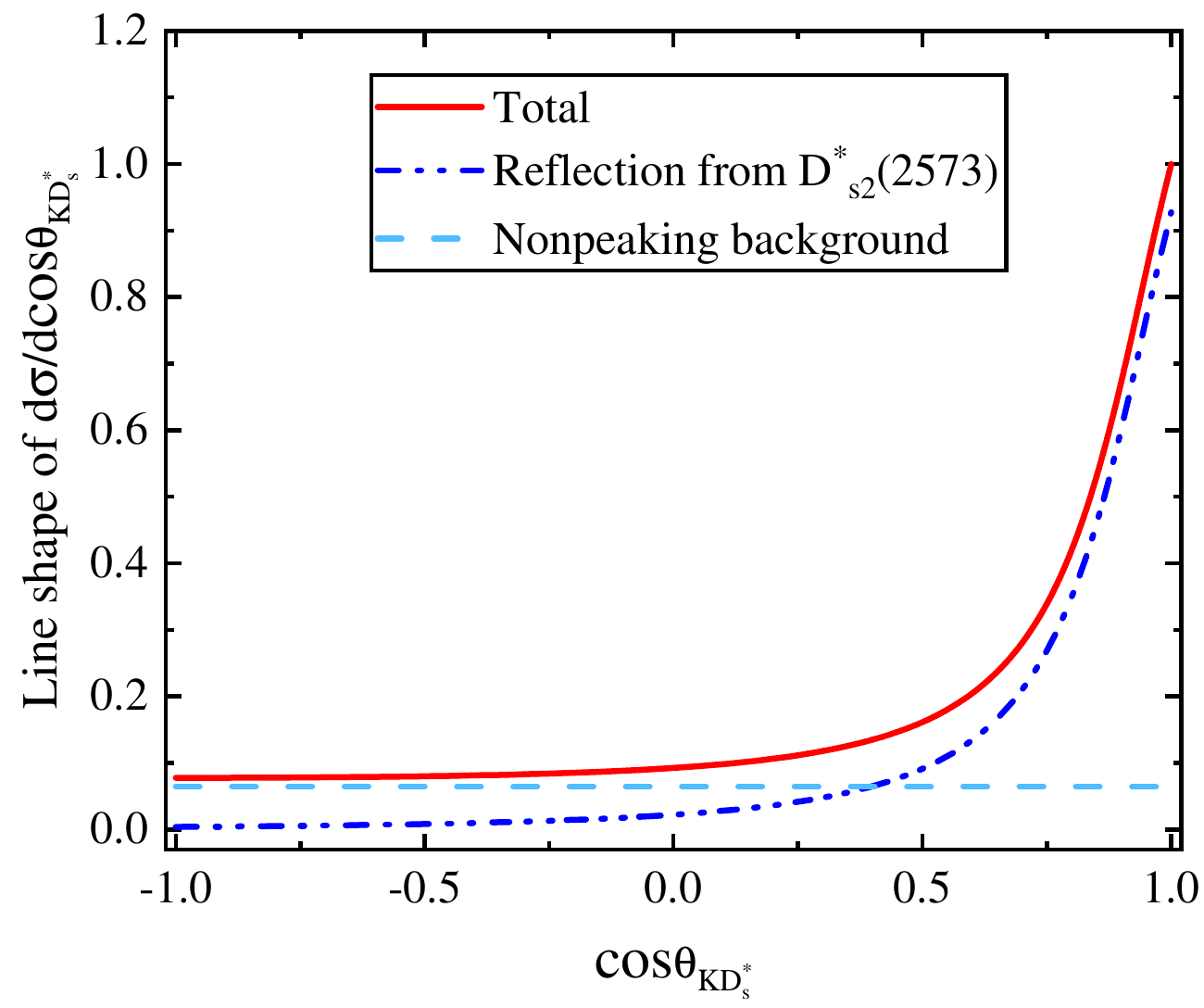}
	\caption{The predicted line shape of differential cross section of  $e^+e^-\to D^0 K^+ D_{s}^{*-}$ vs. cos$\theta_{KD_{s}^{*}}$.}\label{angulardistribution}
\end{figure}

Accumulating data with high statistics, we suggest that the BESIII and BelleII experiments can directly measure the exclusive process of  $e^+e^-\to  D_s^{*}DK$ to clarify the nature of $Z_{cs}(3985)$. In addition to the line shape of the invariant mass spectrum of $D_s^{*}D$, the final states' angular distribution can also reveal some critical information of the dynamical mechanism involved in a process  $e^+e^-\to  D_s^{*}DK$. Here, we define an asymmetry parameter $\mathcal{A}$, which can reflect the asymmetry degree in the angular distribution, i.e.,
\begin{eqnarray}
\mathcal{A}=\frac{\sigma_{\mid cos~\theta_{KD_s^*}\mid>0.5}-\sigma_{\mid \cos~\theta_{KD_s^*}\mid<0.5}}{\sigma_{\mid \cos~\theta_{KD_s^*}\mid>0.5}+\sigma_{\mid cos~\theta_{KD_s^*}\mid<0.5}},
\end{eqnarray}
where $\sigma_{\mid \mathrm{cos}~\theta_{KD_s^*}\mid>0.5}$ is the integrated cross sections of a process $e^+e^-\to  D_s^{*}DK$ in the angle region of $\mid \mathrm{cos}~\theta_{KD_s^*}\mid$ greater than 0.5, and the same is true for $\sigma_{\mid \cos~\theta_{KD_s^*}\mid<0.5}$. $\theta_{KD_s^*}$ corresponds to the angle between a bachelor kaon and the $D_s^*$ direction in the rest frame of the $D_s^{*}D$ system. If the events of $D_s^{*}D$ are from the decay of $Z_{cs}(3985)$ as a tetraquark hadron, the corresponding angular distribution in $\theta_{KD_s^*}$ should be symmetric, where $\mathcal{A}=0$ will be obtained. However, there will be produced an asymmetric angular distribution in our proposed reflection mechanism. In Fig. \ref{angulardistribution}, the line shape of the differential cross section of $e^+e^-\to  D_s^{*-}D^{0}K^+$ vs. $\cos\theta_{KD_s^*}$ is predicted based on the  parameters in Table \ref{parameters}. We can see that there are an obvious increasing distribution for the reflection contribution of $D_{s2}^*(2573)$ from $\mathrm{cos}\theta_{KD_s^*}=-1$ to 1, which corresponds to $\mathcal{A}_{D_{s2}^*(2573)}=0.72$. After considering the non-peaking contributions, the total asymmetric parameter $\mathcal{A}_{\mathrm{Total}}$ is estimated to be around 0.43 and deviates significantly from zero. Thus, the measurement of an asymmetry parameter can be considered as a powerful method to identify the nature of $Z_{cs}(3985)$. These predictions can be left for experimental examination.

In summary, the BESIII collaboration recently observed a new charged charmoiumlike structure $Z_{cs}(3985)^{-}$ in the recoil mass of $K^+$ in a process $e^+e^-\to  D_s^{*-}D^0K^+$ or $D_s^{-}D^{*0}K^+$ at $\sqrt{s}=4.681$ GeV \cite{1830518}. From the above open charm final states, it is easy to conjecture that it contains at least the four-quark component of $c\bar{c}s\bar{u}$ if  $Z_{cs}(3985)^{-}$ is a genuine resonance. Hence, it cannot be a conventional charmonium hadron and must imply a strange essence. In this work, we have proposed a reflection explanation to decode the nature of $Z_{cs}(3985)^{-}$. Our idea is based on the production mechanism of $e^+e^-\to Y(4660) \to D_{s}^{*-}D_{s2}^{*}(2573)^+ \to  D_s^{*-}(D^0K^+)$, where the $P$-wave charmed-strange meson $D_{s2}^{*}(2573)$ from an $S$-wave open-charm decay of  $Y(4660)$ has been found to precisely produce a reflection peak near the threshold of invariant mass of $D_s^{*-}D^0$ at $\sqrt{s}=4.681$ GeV.

 Using a combined fit to experimental data by a line shape on the recoil mass spectrum of $K^+$ at five energy points, we have found that the signal of $Z_{cs}(3985)^{-}$ can indeed be described well by the reflection from $D_{s2}^{*}(2573)$. In addition, because the energy point of $4.681$ GeV just meets the critical relation $\sqrt{s}=m_{D_{s2}^*(2573)}+m_{D_{s}^*}$ of producing a clear reflection peak near the threshold \cite{Wang:2020dmv}, so our proposed mechanism can naturally explain why the obvious signal of $Z_{cs}(3985)^{-}$ is not observed at other CM energy points. Furthermore, we have predicted the angular distribution of an exclusive process $e^+e^-\to  D_s^{*-}D^{0}K^+$ on $\theta_{KD_s^*}$ in our proposed reflection mechanism, which gives an asymmetric parameter $\mathcal{A}_{\mathrm{Total}}=0.43$. This will be an ingenious measurable quantity to identify the nature of the newly observed $Z_{cs}(3985)^{-}$.

The reflection explanation of $Z_{c(s)}$ structures implies that some open charm channels will have a strong coupling with the related $Y$ state, such as  $D^{(*)}D_1(2420)$ for $Y(4220)$ and $D_s^{*}D_{s2}^{*}(2573)$ for $Y(4660)$, to some extent, which should affirm the role of the unquenched effects from open charm channels in understanding the complicated charmoniumlike $Y$ problems again. These hints may be valuable for further revealing the fine internal structure of $Y$ states. We look forward to more experimental data to help us solve the $XYZ$ problem thoroughly in the future.

\section*{ACKNOWLEDGEMENTS}

This work is partly supported by the China National Funds for Distinguished Young Scientists under Grant No. 11825503, the National Program for Support of Top-notch Young Professionals and the 111 Project under Grant No. B20063.

\end{document}